\documentclass[showpacs,twocolumn,prl,aps]{revtex4}
\usepackage{graphicx}

\begin{document}

\title[Short title for running header]{Instability of the chiral d-wave RVB state for the Heisenberg model on triangular lattice and an improvement of the Gutzwiller approximation}
\author{Tao Li}
\affiliation{ Department of Physics, Renmin University of China,
Beijing 100872, P.R.China}
\date{\today}

\begin{abstract}
Through Variational Monte Carlo simulation we show the d-wave RVB
pairing in the Heisenberg model on triangular lattice can be better
described in terms of a two component order parameter. The fully
gapped chiral d-wave RVB state, which is predicted by the mean field
theory to be the unique minimum of variational energy in the two
dimensional representation space of d-wave pairing, is found to be
actually a local maximum and the true minimum of energy is reached
by the non-chiral $d_{xy}$ state with line nodes. We also find that
the usual Gutzwiller approximation, which enjoys great success for
the square lattice system, fails badly on the triangular lattice as
a result of the geometric frustration inherent of the system. An
improved version of the Gutzwiiler approximation is proposed and is
found to give a much better results than the usual one.
\end{abstract}
\maketitle

The search for spin liquid ground state on geometrically frustrated
quantum magnet is a central issue in the study of the strongly
correlated electron systems. On the one hand, the spin liquid state
represents a novel state of matter beyond the Landau-Ginzburg
description. On the other hand, the study of the spin liquid state
is closely related to the study of exotic superconductivity in
strongly correlated electron system. The most famous example in this
respect is the high-T$_{c}$ cuprates. Although the parent compounds
of the cuprates(an ideal quantum antiferromagnet described by the
Heisenberg model on square lattice) exhibit magnetic order, the
study of spin liquid state for the Heisenberg model on square
lattice nevertheless predicts unambiguously the d-wave paring
symmetry for the superconducting order at finite doping and help to
clarify the nature of the pseudogap phase\cite{Lee}.

The discovery of superconductivity in hydrated
$Na_{x}CoO_{2}$\cite{Takada} triggered another round of intensive
researches in this field as a result of the geometric frustration
and the strong electron correlation inherent of the system which has
a triangular lattice\cite{Baskaran,Shastry,Wang,Ogata}.
Historically, it is first on the triangular lattice that the very
concept of spin liquid state was first proposed\cite{Anderson}.
However, after many year's of intensive investigation the nature of
the superconducting state remains illusive. While NMR measurement
suggests a spin singlet pairing with line node, theories of the t-J
model on triangular lattice predicts a chiral d-wave state with full
gap and time reversal symmetry breaking. In recent years, the
possibility of spin liquid state and exotic superconducting
state(with self-doped charge carrier) in half-filled systems with an
anisotropic triangular lattice and reduced strength of local
correlation also arose great interest in the field\cite{Kanoda}.
Here the same uncertainty on the nature of the spin liquid state
exist\cite{LeeTK,Motrunich,Sorella,Gros}. While some theory predicts
a spin liquid state with a open Fermi surface for spin excitation,
others argue spin pairing is unavoidable in the spin liquid state.

As compared to the square lattice, the triangular lattice is more
complex in that there are more choices for the spin pairing pattern
on the triangular lattice. According to group theory, spin pairing
between neighboring sites on the square lattice can be either
extended s-wave or d-wave, both of which belong to a one dimensional
irreducible representation of the $C_{4v}$ group. However, on the
triangular lattice with a point group of $C_{6v}$, the d-wave
pairing channel becomes two dimensional and it is a non-trial
problem to decide what kind of pairing pattern is actually realized
within the two dimensional representation space of the d-wave
pairing(the extended s-wave pairing is in general less stable than
d-wave pairing for system with strong local correlation).

The slave Boson mean field theory and the Gutzwiller approximation
based on it is widely adopted to study the problem of spin liquid
and the pairing symmetry of the strongly correlated
system\cite{Lee}. Such a approach receives great succuss on the
square lattice. For example, it predicts correctly the d-wave
pairing symmetry of the high-T$_{c}$ cuprates\cite{Kotliar}. With
the Gutzwiller approximation, the mean field theory can even give a
quantitative correct estimation of the variational energy with an
error less than 5 percent\cite{Zhang,Hsu}. On the triangular
lattice, the slave boson mean field theory predicts unambiguously
that the chiral d-wave pairing with time reversal symmetry breaking
and a full gap is the most stable in two dimensional representation
space and its condensation energy is more than 30 percent lower than
that of the non-chiral(real) d-wave pairing state. The chiral d-wave
state is widely believed to be the best choice for the triangular
lattice.

In this paper, we show the usual Gutzwiller approximation fails
badly for the Heisenberg model on the triangular lattice. It
overestimates the condensation energy by more than 50 percent and
predicts incorrect order for the relative stability of the chiral
d-wave and non-chiral d-wave state. Through direct variational Monte
Carlo simulation, we find the chiral d-wave RVB state is actually a
local maximum in the two dimensional representation space of d-wave
pairing and the true minimum of variational energy is reached by the
non-chiral $d_{xy}$ state with line nodes. The anisotropy of the
variational energy in the representation space is found to be much
smaller than that predicted by the mean field theory and the RVB
pairing should be more appropriately described in terms of a two
component order parameter. We also find the failure of the usual
Gutzwiller approximation on triangular lattice can be attributed to
the geometric frustration inherent of the system. Based on this
understanding, we propose an improved Gutzwiller approximation
scheme and find it works much better than the usual one.

The Heisenberg model under consideration reads
\begin{equation}
\mathrm{H}=J\sum_{<i,j>}\vec{\mathrm{S}}_{i}\cdot\vec{\mathrm{S}}_{j},
\end{equation}
here $\sum_{<i,j>}$ means sum over nearest neighbors on the
triangular lattice. The slave Boson mean field theory of the RVB
state is established by introducing the slave particle
representation of the spin operator
$\vec{S}_{i}=\frac{1}{2}f^{\dagger}_{i,\alpha}\vec{\sigma}_{\alpha,\beta}
f_{i,\beta}$ and then use the mean field order parameter
$\chi_{ij}=\langle f^{\dagger}_{i,\alpha}f_{j,\alpha}\rangle$ and
$\Delta_{ij}=\langle\epsilon_{\alpha,\beta}f^{\dagger}_{i,\alpha}f^{\dagger}_{j,\beta}\rangle$
to decouple the Hamiltonian written in terms of the slave particles.
The slave particles must be subjected to the no double occupancy
constraint to represent the spin operator faithfully. In mean field
treatment, such local constraint is relaxed to a constraint on the
average particle number on each site.

Assuming translational invariance for the RVB order parameter
$\chi_{i,j}$ and $\Delta_{i,j}$, the mean field Hamiltonian for the
slave particle has the form,
\begin{equation}
H_{MF}=\sum_{k,\alpha}\xi_{k}f^{\dagger}_{k,\alpha}f_{k,\alpha}
       +\sum_{k}(\Delta_{k}f^{\dagger}_{k,\uparrow}f^{\dagger}_{-k,\downarrow}+h.c.)
\end{equation}
whose ground state reads
\begin{equation}
|BCS\rangle=\prod_{k}(u_{k}+v_{k}f^{\dagger}_{k\uparrow}f^{\dagger}_{-k\downarrow})|0\rangle,
\end{equation}
in which $\frac{v_{k}}{u_{k}}=\frac{\Delta_{k}}{\xi_{k}+E_{k}}$,
$E_{k}=\sqrt{\xi_{k}^{2}+\Delta_{k}^{2}}$. Instead of determining
$\chi_{i,j}$ and $\Delta_{i,j}$ from the self-consistent equation,
we will take the wave function Eq.(3) as a variational description
of the system at the mean field level. For simplicity, $\chi_{i,j}$
will be assumed to be a real constant and be nonzero for nearest
neighboring sites only. The hopping energy $\xi_{k}$ then reads
\begin{equation}
\xi_{\mathbf{k}}=-2(\cos
k_{x}+2\cos\frac{k_{x}}{2}\cos\frac{\sqrt{3}k_{y}}{2})-\mu,
\end{equation}
here we have set the hopping integral as one. The chemical potential
is determined by the mean field equation for the particle number and
is not treated as an independent variational parameter. Thus the
magnitude of the gap function, $\Delta$, is the only variational
parameter in the theory.

The physical wave function of the spin system should satisfy the no
double occupancy. This can be achieved by taking the Gutzwiller
projection on the mean field wave function. The Gutzwiller projected
mean field state takes the form
\begin{equation}
|\Psi\rangle=P_{\mathrm{G}}\left(\sum_{i,j}a_{i-j}c^{\dagger}_{i,\uparrow}c^{\dagger}_{j,\downarrow}\right)^{\frac{N_{e}}{2}}|0\rangle,
\end{equation}
where the Cooper pair wave function $a_{i-j}$ is given by
\begin{equation}
a_{i-j}=\sum_{\mathbf{k}}\frac{\Delta_{\mathbf{k}}}{\epsilon_{\mathbf{k}}+\sqrt{\epsilon_{\mathbf{k}}^{2}+|\Delta_{\mathbf{k}}|^{2}}}\exp^{i\mathbf{k}(\mathbf{R}_{i}-\mathbf{R}_{j})},
\end{equation}
and $P_{\mathrm{G}}$ is the projection into the subspace of no
double occupancy. The Gutzwiiler projected wave function is usually
treated by the variational Monte Carlo simulation method as
analytical calculation is difficult for it. An estimate of
variational energy in the Gutzwiller projected wave function can be
given by the so called Gutzwiller approximation, in which the
expectation value of a physical observable in the Gutzwiller
projected state is approximated by its expectation value in the mean
field state multiplied with a Gutzwiller factor. For the Heisenberg
system,  the Gutzwiller factor for spin correlation is usually taken
to be $g_{s}=4$.

Now we determine the structure of the pairing order parameter. Here
we will consider the d-wave pairing only. On the triangular lattice,
the d-wave pairing belongs to a two dimensional irreducible
representation of the point group of $C_{6v}$ and is thus more
complex than its counterpart on the square lattice, which has a
unique basis function of the form of $d_{x^{2}-y^{2}}$. The two
basis functions of this two dimensional representation have the form
of $d_{x^{2}-y^{2}}$ and $d_{xy}$. The symmetry of the system
guarantee the invariance of the free energy under rotation in the
two dimensional representation space at the quadratic level. Any
anisotropy in the representation space must be attributed to higher
order terms in the Landau expansion. Thus, if such anisotropy is
small, one should better describe the pairing state with a
multi-component order parameter.

\begin{figure}[h!]
\includegraphics[width=8cm,angle=0]{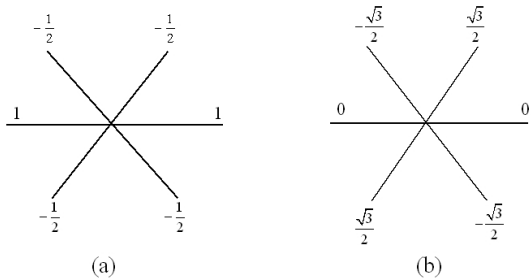}
\caption{The two basis functions of d-wave pairing on triangular
lattice.(a)$d_{x^{2}-y^{2}}$,(b)$d_{xy}$.} \label{fig1}
\end{figure}

In momentum space, these two basis functions of the d-wave
representation have the form of $\Phi_{1}(\mathbf{k})=\cos
k_{x}-\cos \frac{k_{x}}{2}\cos \frac{\sqrt{3}k_{x}}{2}$ and
$\Phi_{2}(\mathbf{k})=\sqrt{3}\sin \frac{k_{x}}{2}\sin
\frac{\sqrt{3}k_{x}}{2}$. In real space, they behave as $\cos
2\varphi$ and $\sin 2\varphi$ respectively(see Fig.1), where
$\varphi$ is the angle the bond made with $x-$axis. The general
order parameter is given by
$\Delta_{\mathbf{k}}=\Delta[\eta_{1}\Phi_{1}(\mathbf{k})+\eta_{2}\Phi_{2}(\mathbf{k})]$,
where $\eta_{1}$ and $\eta_{2}$ are two complex numbers satisfying
$|\eta_{1}|^{2}+|\eta_{2}|^{2}=1$. Up to a global phase, the gap
function can then be parameterized as follows
\begin{equation}
\Delta_{\mathbf{k}}=\Delta[\cos\theta\Phi_{1}(\mathbf{k})+\sin\theta
\exp(i\phi)\Phi_{2}(\mathbf{k}))],
\end{equation}
in which $\Delta,\theta$ and $\phi$ are all real numbers. In this
parameterizition, the chiral d-wave state is given by
$(\theta,\phi)=(\frac{\pi}{4},\frac{\pi}{2})$. Owing to the gauge
symmetry and symmetry of space inversion and time reversal, it is
easy to see that we only need to consider the value of $\theta$ and
$\phi$ in the region $(\theta,\phi)\in [0,\pi/2]\times[0,\pi/2]$.

At the mean field level, the chiral $d_{x^{2}-y^{2}}+id_{xy}$ state
gives the lowest energy in this space while the maximum in energy is
reached by the non-chiral $d_{xy}$ state. Note also that the energy
is periodic in $\theta$ when $\phi=0$ with a period of $\pi/3$, as a
result of the six-fold rotational symmetry of the triangular
lattice. A full scan of the variational energy(optimized with
respect to the parameter $\Delta$) estimated from the usual
Gutzwiller approximation with $g_{s}=4$ is shown in Fig.2. For any
given $\theta$, a real gap function always gives higher variational
energy than a complex gap function.

\begin{figure}[h!]
\includegraphics[width=9cm,angle=0]{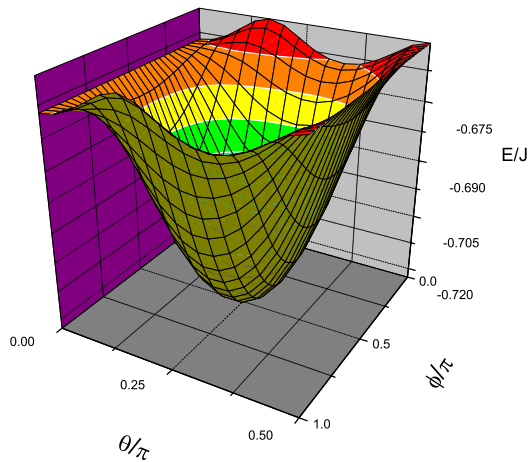}
\caption{A full scan of the variational energy(optimized with
respect to $\Delta$)calculated from the Gutzwiller approximation
with $g_{s}=4$ in the representation space of d-wave pairing.}
\label{fig2}
\end{figure}

Figure 3 provides a more quantitative measure on the anisotropy in
the representation space at the mean field level, in which the
variational energy calculated from the Gutzwiller approximation with
$g_{s}=4$ is compared for the chiral $d_{x^{2}-y^{2}}+id_{xy}$ state
and the non-chiral $d_{xy}$ state as a function of $\Delta$. It is
found that the chiral d-wave state enjoys a more than 30 percent
lower condensation energy than its non-chiral counterpart. For this
reason, the two component nature of the order parameter is
irrelevant and we can take the chiral d-wave state as the unique
choice for spin pairing.

\begin{figure}[h!]
\includegraphics[width=8cm,angle=0]{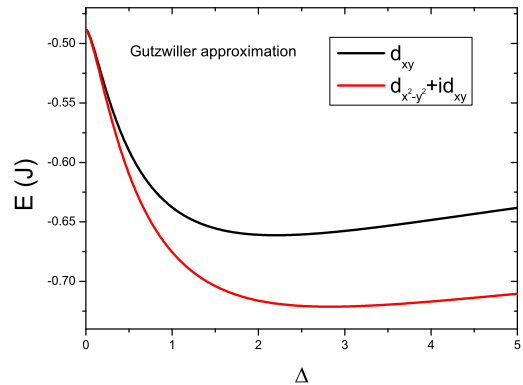}
\caption{The variational energy as a function of $\Delta$ calculated
from the Gutzwiller approximation with $g_{s}=4$ for the chiral
$d_{x^{2}-y^{2}}+id_{xy}$ state and the non-chiral $d_{xy}$ state.}
\label{fig3}
\end{figure}

Now we present the results of Variational Monte Carlo simulation.
The simulation is done on a $14\times 14$ lattice with
periodic-antiperiodic boundary condition. $10^{7}$ statistically
independent samples are used to estimate the variational energy. The
result for the chiral $d_{x^{2}-y^{2}}+id_{xy}$ state and the
non-chiral $d_{xy}$ state are shown in Figure 4. The result differs
from the mean field result in three important aspects. Firstly, the
order of relative stability between the chiral
$d_{x^{2}-y^{2}}+id_{xy}$ state and the non-chiral $d_{xy}$ state is
now reversed. Secondly, the difference in condensation energy is
much smaller than that predicted by the mean field theory, being
below 3 percent. Lastly, the variational energy estimated from the
Gutzwiller approximation is about 50 percent lower than that
calculated directly from the Monte Carlo simulation. We thus
conclude that the usual Gutzwiller approximation fails badly on the
triangular lattice.
\begin{figure}[h!]
\includegraphics[width=8cm,angle=0]{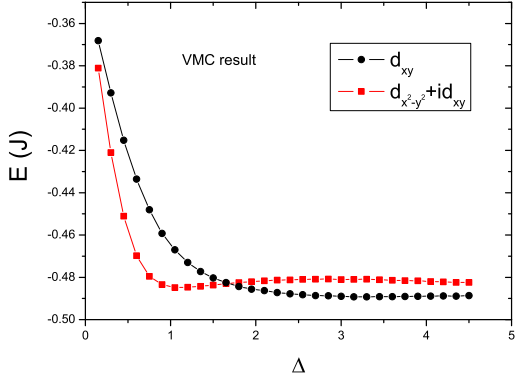}
\caption{The variational energy as a function of $\Delta$ for the
chiral $d_{x^{2}-y^{2}}+id_{xy}$ state and the non-chiral $d_{xy}$
state calculated from VMC simulation. The simulation is done on a
$14\times14$ lattice with periodic-antiperiodic boundary condition.}
\label{fig4}
\end{figure}

\begin{figure}[h!]
\includegraphics[width=9cm,angle=0]{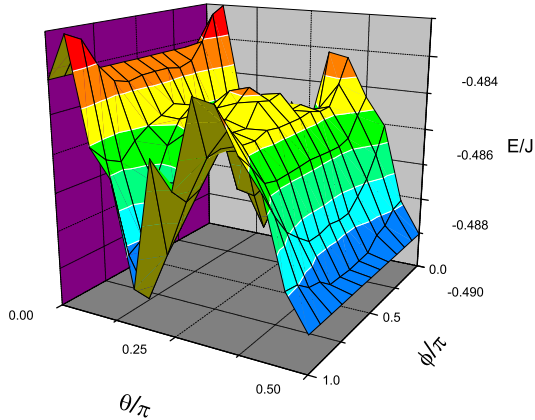}
\caption{A full scan of the variational energy(optimized with
respect to $\Delta$) in the representation space of d-wave pairing
calculated from VMC simulation. The simulation is done on a
$16\times16$ lattice with periodic-antiperiodic boundary condition.}
\label{fig5}
\end{figure}

A full scan of the variational energy(optimized with respect to
$\Delta$) as a function of $\theta$ and $\phi$ is shown in Figure 5.
The simulation is done on a $16\times16$ lattice with
periodic-antiperiodic boundary condition. The variational energy is
approximately periodic in $\theta$ with a period of $\pi/3$ when
$\phi=0$, as required by the six-fold rotational symmetry of the
triangular lattice. The small deviation from such a periodicity,
especially the small extra peaks around $(\theta,\phi)=(\pi/12,0)$
and $(\theta,\phi)=(5\pi/12,0)$, are caused by finite size effect
which is the strongest for the non-chiral state with line node. From
the Figure we see the true minimum of the variational energy is
reached by the non-chiral $d_{xy}$ state, while the chiral
$d_{x^{2}-y^{2}}+id_{xy}$ state now becomes a local maximum. At the
same time, the anisotropy of the condensation energy in the
representation space, being less than 5 percent, is much smaller
than that predicted by the mean field theory. For this reason, it is
more appropriate to describe to the d-wave pairing in terms of a two
component order parameter.

\begin{figure}[h!]
\includegraphics[width=8cm,angle=0]{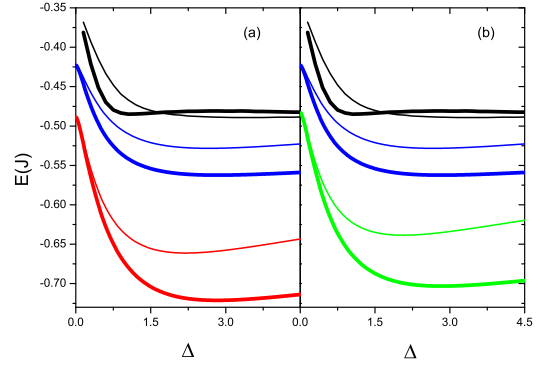}
\caption{The variational energy of the chiral
$d_{x^{2}-y^{2}}+id_{xy}$ state(thick lines) and the non-chiral
$d_{xy}$ state(thin lines) as a function of $\Delta$ calculated from
VMC simulation(black lines), one site Gutzwiller approximation(red
lines), two site Gutzwiller approximation(green lines) and the three
site Gutzwiller approximation(blue lines).} \label{fig6}
\end{figure}

Then, why does the usual Gutzwiller approximation, which perform
well for the square lattice, fails so badly on the triangular
lattice? To understand this, we rewrite the expectation value
$\langle s_{i}^{z}s_{j}^{z}\rangle$ in the Gutzwiller projected
state as follows,
\begin{equation}
\langle
s_{i}^{z}s_{j}^{z}\rangle=\sum_{\alpha}W_{\alpha}s_{i}^{z}(\alpha)s_{j}^{z}(\alpha)
=\sum_{\alpha}W_{\alpha}^{0}\frac{W_{\alpha}}{W_{\alpha}^{0}}s_{i}^{z}(\alpha)s_{j}^{z}(\alpha),
\end{equation}
in which $\alpha$ denotes one of the four spin
configurations($\uparrow\uparrow,\uparrow\downarrow,\downarrow\uparrow,\downarrow\downarrow$)
on site $i$ and $j$, $W_{\alpha}$ and $W_{\alpha}^{0}$ denote the
probabilities for the spin configuration $\alpha$ to appear in the
projected and mean field state. If we neglect the $\alpha$
dependence of the ratio $\frac{W_{\alpha}}{W_{\alpha}^{0}}$, we get
\begin{equation}
\langle s_{i}^{z}s_{j}^{z}\rangle \approx g_{s}
\sum_{\alpha}W_{\alpha}^{0}s_{i}^{z}(\alpha)s_{j}^{z}(\alpha)=g_{s}\langle
s_{i}^{z}s_{j}^{z}\rangle_{0},
\end{equation}
in which $g_{s}=\frac{1}{\sum_{\alpha}W_{\alpha}^{0}}$, $\langle
\rangle_{0}$ denotes the expectation value in the mean field state.
We note that such a approximation scheme becomes exact if
$\frac{W_{\alpha}}{W_{\alpha}^{0}}$ is indeed $\alpha$ independent.
In the BCS mean field state studied in this paper, the Gutzwiller
factor $g$ is easily calculated to be given by
$g_{s}=\frac{4}{1+|\chi_{ij}|^{2}+|\Delta_{ij}|^2}$, where
$\chi_{ij}=\sum_{\sigma}\langle c^{\dagger}_{i,\sigma}c_{j,\sigma}
\rangle_{0}$,
$\Delta_{ij}=\sum_{\sigma}\epsilon_{\sigma,\bar{\sigma}}\langle
c^{\dagger}_{i,\sigma}c^{\dagger}_{j,\bar{\sigma}} \rangle_{0}$.
This is nothing but the two site version of the(improved) Gutzwiiler
approximation derived by Hsu\cite{Hsu}. The usual Gutzwiller
approximation with $g_{s}=4$ can be derived by simply neglecting the
inter-site correlation and will be called the single site version of
Gutzwiller approximation in the following. In Figure 6, we show the
variational energy calculated from both the two-site and the single
site version of Gutzwiiler approximation and compared them to the
result of VMC simulation. Although the two-site Gutzwiller
approximation improve the result of the single site Gutzwiller
approximation, none of them is truly satisfactory.

The failure of the above approximation schemes can be attributed to
the assumption that the ratio $\frac{W_{\alpha}}{W_{\alpha}^{0}}$
being $\alpha$ independent. On the triangular lattice, a given pair
of nearest neighboring sites $i$ and $j$ are neighbored by a third
site $k$. In the mean field state, the ratio between the
probabilities for parallel and antiparallel spin alignment on site
$i$ and $j$, namely
$R=\frac{W_{\uparrow\downarrow}^{0}}{W_{\uparrow\uparrow}^{0}}$,
depends crucially on the occupation of their common neighbor of site
$k$. For example, if site $k$ is empty, then site $i$ and $j$ can
take the full advantage of forming singlet pair without been
frustrated by the spin on site $k$. Similarly, if site $k$ is doubly
occupied, then two antiparallel spins on site $i$ and $j$ can each
form singlet pair with the electron on site $k$. On the other hand,
if site $k$ is singly occupied, no matter what the spin it has, its
coupling to the two antiparallel spins on site $i$ and $j$ is
frustrated. For this reason, the ratio $R$ is larger when site $k$
is either empty or doubly occupied than it is singly occupied. This
explains the breakdown of the assumption that
$\frac{W_{\alpha}}{W_{\alpha}^{0}}$ being $\alpha$ independent. To
substantiate these arguments, we plot in Figure 7 the ratio $R$ for
the chiral $d_{x^{2}-y^{2}}+id_{xy}$ state as a function of $\Delta$
when site $k$ is either empty($R_{e}$), doubly occupied($R_{d}$), or
singly occupied($R_{s}$). From the figure we see $R_{e}$ is always
higher than $R_{s}$ and $R_{d}$ grows much faster than $R_{s}$ and
exceeds it at large $\Delta$. All these observations are consistent
with the qualitative arguments raised above.

The above reasoning also suggest a way to improve the Gutzwiller
approximation. To reach this goal, we simply extend the two site
version of the approximation to a three site version and use
$\alpha$ to denotes the eight possible spin configurations on a
elementary triangle of the lattice. In this scheme, the variational
energy per triangle reads
\begin{equation}
\langle
(s_{i}^{z}s_{j}^{z}+s_{i}^{z}s_{k}^{z}+s_{j}^{z}s_{k}^{z})\rangle=g_{s}\langle(s_{i}^{z}s_{j}^{z}+s_{i}^{z}s_{k}^{z}+s_{j}^{z}s_{k}^{z})\rangle_{0},
\end{equation}
in which $g_{s}=\frac{1}{\sum_{\alpha}W_{\alpha}^{0}}$, $i,j,k$
denotes the three neighboring sites on an elementary triangle. The
calculation of $W_{\alpha}^{0}$ is straightforward but tedious and
we will not present analytical expression for it here. For
illustrative purpose, we show here the calculation of
$W_{\alpha}^{0}$ for one of the eight spin configuration, namely the
configuration with all three spins aligned up. It is given by
\begin{equation}
W_{\uparrow\uparrow\uparrow}^{0}=\langle(1-n_{i\downarrow})n_{i\uparrow}(1-n_{j\downarrow})n_{j\uparrow}(1-n_{k\downarrow})n_{k\uparrow}
\rangle_{0},
\end{equation}
which can be evaluated with the Wick expansion. The expression for
other $W_{\alpha}^{0}$ can be similarly constructed and calculated.

\begin{figure}[h!]
\includegraphics[width=8cm,angle=0]{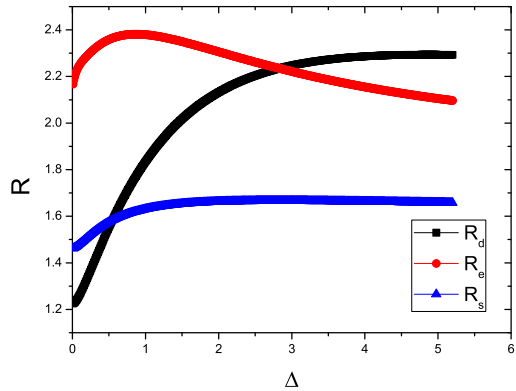}
\caption{The ratio between the probabilities for antiparallel and
parallel spin alignment on a pair of nearest neighboring sites $i$
and $j$ on the triangular lattice when their common neighbor site
$k$ is doubly occupied($R_{d}$), empty($R_{e}$) and singly
occupied($R_{s}$). The calculation is done for the chiral
$d_{x^{2}-y^{2}}+id_{xy}$ state.} \label{fig7}
\end{figure}

In Figure 6, we compare the variational energy calculated from the
one site, two site and the three site version of the Gutzwiiler
approximation with the VMC results. While the two site version of
Gutzwiller approximation is seen to have only a very limited
improvement over the one site approximation, the three site
approximation provides a much more substantial improvement over the
two. We thus conclude that the geometric frustration is indeed at
the root of the failure of the usual Gutzwiller approximation on the
triangular lattice system. It should also be noted that the order of
relative stability between the chiral and the non-chiral d-wave
state is still incorrect at the three site version of the Gutzwiller
approximation and higher order of approximation is needed to recover
the correct order.

In conclusion, we have studied the problem of RVB pairing symmetry
for the Heisenberg model on the triangular lattice, which is
relevant for both the study of nature of the Mott insulating state
of the half filled Hubbard model on the triangular lattice and the
superconducting state in the doped system. Unlike the square lattice
system, the d-wave pairing on the triangular lattice require a
description in terms of a two component order parameter. Contrary to
the previous study based on usual Gutzwiiler approximation, we find
the chiral-d wave state is actually a local maximum rather than
local minimum in the two dimensional representation space of the
d-wave pairing. We also find the anisotropy of condensation energy
in the representation space is very small(less than 5 percent) and
the true minimum is reached by the non-chiral $d_{xy}$ with line
nodes. We find the usual Gutzwiller approximation, which perform
well on square lattice, fails badly on the triangular lattice. The
failure of the usual Gutzwiller approximation scheme is traced back
to the geometric frustration inherent of the triangular lattice and
an improved version of the Gutzwiller approximation is proposed
based on this understanding and is found to result in substantial
improvement over the usual approximation scheme.

The author is supported by NSFC Grant No. 10774187 and National
Basic Research Program of China No. 2007CB925001 and Beijing Talent
Program.

\end{document}